\documentclass[pra,preprint,amsmath,amssymb]{revtex4}
\usepackage{bm}
\usepackage{graphics}
\newcommand{\dg}{\dagger}
\begin{document}

\title{Quantum instabilities in the system of identical  bosons }
\author{Valery S. Shchesnovich}
\email{valery@loqnl.ufal.br}
\affiliation{ Departamento de
F\'{\i}sica - Universidade Federal de Alagoas, Macei\'o AL
57072-970, Brazil }

\begin{abstract}
The quantum instability of the mean-field theory for identical
bosons  is shown to be described by an appropriate Bogoliubov
transformation. A connection between the quantum and classical
linear stability theories is indicated. It is argued that the
instability rate in a system of identical bosons must be strongly
affected by the nonlinear terms (interactions). In the case of the
repulsive interactions or strong attractive interactions the
instability rate is suppressed. On the other hand, a weak
attraction  significantly enhances the instability rate. The
results can have applications in the field of Bose-Einstein
condensates of dilute quantum gases.
\end{abstract}

\maketitle

Under some conditions, a quadratic Hamiltonian describing a system of identical
bosons can be diagonalized by the Bogoliubov transformation \cite{BGLB}. The
physical sense of the diagonalization lies in a redefinition of the vacuum state
and introduction of quasiparticles satisfying the boson commutation relations. The
diagonalization is used, for instance, in description of the excitations in a
Bose-Einstein condensate of  dilute quantum gas \cite{BEC}.

Diagonalization is not always possible which means that the quasiparticle
excitations may give rise to an instability \cite{Anglin}. Below it is shown that
such instabilities are described by an appropriate Bogoliubov transformation which,
however, is not related to the diagonalization procedure. Moreover, it is argued
that even weak interactions (i.e. the higher-order terms) can cause a dramatic
effect on the instability rate.

Driving a  quantum many-body system through an instability region by changing the
system parameters is an important tool in the quantum state engineering in BEC
physics \cite{BEC}, for instance,  in the process of the atomic association
\cite{ExpAtomAssoc1,ExpAtomAssoc2,TheorAtomAssoc1} and molecular dissociation in
BEC \cite{QEffAtomMol,QCurveCross2}. In a region of instability the mean-field
theory breaks down  \cite{TheorAtomAssoc1,QEffAtomMol,QCorrMFT}. The instability in
system of identical bosons is also referred to as the curve crossing
\cite{QCurveCross2,QCurveCross1} and is related to the Landau-Zener theory
\cite{QCurveCross2,Anglin}. In a condensate with the attractive interactions the
quantum corrections are crucial for description of the dynamics when  the mean
field state is unstable \cite{BSB}.

Let us first  briefly recall the basics of the diagonalization procedure
\cite{BGLB}. Consider a Hamiltonian quadratic in the boson creation ($b_j^\dg$) and
annihilation ($b_j$) operators:
\begin{equation}
H = \sum_{k,l}A_{k,l} b^\dg_k b_l + \frac{1}{2}\sum_{k,l}B_{k,l}
b^\dg_k b^\dg_l + \frac{1}{2}\sum_{k,l}B^*_{k,l} b_k b_l,
\label{EQ1}\end{equation}
where $A^\dg = A$ and $\tilde{B} = B$ (here $\dg$,
$\;{\null}_{\widetilde{}}\;$, and $*$ denote the Hermitian
conjugation, the transposition, and the complex conjugation,
respectively). The diagonalization of the  Hamiltonian (\ref{EQ1})
introduces new boson operators, say $\xi^\dg_j$ and $\xi_k$, such
that the corresponding matrix $A$ becomes diagonal and $B$
vanishes.

The fact that the new operators satisfy the boson commutation
relations  together with the linearity and homogeneity of the
transformation gives immediately the most general form of the
Bogoliubov transformation:
\begin{equation}
\left(\begin{array}{c}\xi \\ \xi^\dg\end{array}\right)
=\left(\begin{array}{cc} U^\dg  & -V^\dg  \\
-\tilde{V}  & \tilde{U}  \end{array}\right)\left(\begin{array}{c}b \\
b^\dg\end{array}\right), \quad U^\dg  U  - V^\dg  V  = I,\quad
\tilde{U} V  - \tilde{V} U = 0.
\label{EQ2}\end{equation}
Here, for simplicity, the index-free operators denote the whole
column of such operators, e.g. $\xi \equiv (\xi_1,\ldots,\xi_n)$.
Note that the conditions on matrices $U$ and $V$ in equation
(\ref{EQ2}) also guarantee the existence of the inverse
transformation
\begin{equation}
\left(\begin{array}{c}b \\
b^\dg\end{array}\right)
=\left(\begin{array}{cc} U  & V^*  \\
V  & U^*  \end{array}\right)\left(\begin{array}{c}\xi \\
\xi^\dg\end{array}\right).
\label{EQ3}\end{equation}

To diagonalize the Hamiltonian (\ref{EQ1}) the Bogoliubov
transformation must be such that columns of the transformation in
equation (\ref{EQ3}) are the eigenvectors of the following matrix
eigenvalue problem
\begin{equation}
\left(\begin{array}{cc}I & 0 \\
0 & -I\end{array}\right)
\left(\begin{array}{cc} A & B \\
B^* & A^* \end{array}\right)\left(\begin{array}{c}U_{\cdot,k} \\
V_{\cdot,k}\end{array}\right) = \lambda_k\left(\begin{array}{c}U_{\cdot,k} \\
V_{\cdot,k}\end{array}\right),
\label{EQ4}\end{equation}
where $U_{\cdot,k}$ and $V_{\cdot,k}$ denote the $k$-th columns of the respective
matrices. If the diagonalization is possible, the Hamiltonian (\ref{EQ1}) in the
quasiparticle representation reads $H = \sum \lambda_k \xi^\dg_k\xi_k - \sum
\lambda_k\left(\sum_l |v_{l,k}|^2\right)$, where $v_{l,k}$ are the elements of the
matrix $V$ \cite{BGLB}. Note that the number of quasiparticles $N = \sum
\xi^\dg_k\xi_k$ is conserved.

The diagonalization procedure described above is closely related to  the classical
theory of stability of the stationary points of the Hamiltonian systems
\cite{MacKay}. Indeed, the second matrix on the l.h.s. of equation (\ref{EQ4}),
\begin{equation}
\mathcal{H} =
\left(\begin{array}{cc} A & B \\
B^* & A^* \end{array}\right),
\label{EQ5}\end{equation}
is the Hessian  of the  classical Hamiltonian system corresponding to the quantum
Hamiltonian (\ref{EQ1}) when the boson operators $b$ and $b^\dg$  are replaced by
the $c$-numbers. Hence, the eigenvalues $\lambda_k$ of the eigenvalue problem
(\ref{EQ4}) come in quartets $\{\lambda,\lambda^*,-\lambda,-\lambda^*\}$.  For the
diagonalization to exist it is necessary that  there are $n$ non-zero real
eigenvalues and  $n$ corresponding eigenvectors with the positive normalization,
i.e. $U^\dg  U  - V^\dg V = I$.

The simplest  sufficient condition for the diagonalization is the positivity of the
corresponding Hessian. In the classical stability theory the quadratic form of the
Hessian is the first non-zero term in the energy expansion, thus its positivity
implies linear stability \cite{MacKay}. The physical sense of the positivity of the
Hessian is that the linearized system about the stationary solution is equivalent
to a system of uncoupled oscillators (in the stationary case). In the quantum case
the corresponding Hamiltonian, quadratic in boson operators, can be diagonalized.
However, it is clear that the diagonalization is not always possible (for example,
the Hamiltoian (\ref{EQ6}) below cannot be diagonalized when $|\delta|<\gamma$). In
this case there no interpretation in terms of the uncoupled oscillators since the
linearized system is hyperbolic.

A mean-field theory for bosons can be considered stable if the
quasiparticle excitations remain bounded (for instance, when the
Hessian is positive). There is a correspondence between the
classical and quantum stability. Physically such a correspondence
seems to be quite clear: due to the linearity of the Heisenberg
evolution equations (for $b$ and $b^\dg$) the stability in the
classical sense implies  that the diagonalization of the
corresponding Hamiltonian, quadratic in boson operators, is
possible. In the classical theory it is established that under a
variation of the parameters the solution to a Hamiltonian system
may become unstable due to collision of the imaginary eigenvalues
with the energy of opposite signs (the opposite Krein signatures)
or due to a collision of the eigenvalues at zero \cite{MacKay}.
Hence, the simplest quantum instability is due to a collision of
two frequencies at a resonance.

The simplest example of the quantum instability (or non-diagonalizable boson
Hamiltonian)  has already appeared in the description of a boson system driven
through a resonance \cite{QCurveCross1,QCurveCross2,Anglin}. The corresponding
Hamiltonian reads \cite{QCurveCross2,Anglin}
\begin{equation}
H = [\omega + \delta(t)]a_1^\dg a_1 + [-\omega + \delta(t)]a_2^\dg
a_2 + \gamma(a_1^\dg a_2^\dg + a_1a_2).
\label{EQ6}\end{equation}
This Hamiltonian  conserves the difference of the number of
bosons: $[H,a_1^\dg a_1 - a_2^\dg a_2] = 0$. The total number of
bosons is not conserved which is a reflection of the fact that the
Hamiltonian (\ref{EQ6}) appears as a linearization of the full
nonlinear Hamiltonian about the mean-field state or in the case of
a macroscopically populated  source of bosons. For instance, such
a Hamiltonian appears in the description of the formation of
atomic pairs  by the dissociation of a molecular condensate
\cite{QCurveCross1,QCurveCross2}. In the case of one highly
populated molecular mode, while the atomic mode populations are
not large, the model Hamiltonian is $H_{am} = [\omega +
\delta(t)]\psi_{1}^\dg \psi_{1} + [-\omega +
\delta(t)]\psi_{2}^\dg \psi_{2}  + (g\Psi_m\psi_{1}^\dg
\psi^\dag_{1} + g^*\Psi_m^\dag\psi_{1}\psi_{2} )$. If a
macroscopic number of molecules  does not change considerably, the
annihilation operator for the molecular state can be approximated
by an order parameter, $\Psi_m \to \langle\Psi_m\rangle$. Defining
$\gamma = |g\langle\Psi_m\rangle|$ and $\psi_{j} = e^{i\chi}a_j$
with $\chi = \mathrm{arg}\{g\langle\Psi_m\rangle\}/2$ we obtain
the Hamiltonian (\ref{EQ6}).

For the Hamiltonian (\ref{EQ6}) the corresponding Hessian reads
\begin{equation}
\mathcal{H} = \left(\begin{array}{cc} A & B \\
B^* & A^* \end{array}\right) =  \left(\begin{array}{cc}
\begin{array}{cc} \delta + \omega & 0 \\ 0 & \delta -\omega
\end{array} & \begin{array}{cc} \quad 0& \quad\gamma \quad\\ \quad\gamma & \quad 0\quad
\end{array}\\
\begin{array}{cc} \quad 0& \quad\gamma \quad\\ \quad\gamma & \quad 0\quad
\end{array} & \begin{array}{cc} \delta + \omega & 0 \\ 0 & \delta -\omega
\end{array}  \end{array}\right).
\label{EQ7}\end{equation}
The characteristic equation is bi-quadratic: $\lambda^4 -
2(\omega^2 +\delta^2 - \gamma^2)\lambda^2 +(\gamma^2 -
\delta^2+\omega^2)^2 = 0$, it has the following four roots
$\lambda = \pm (\omega \pm \sqrt{\delta^2 - \gamma^2})$ (the two
$\pm$ signs can be arbitrarily selected). Therefore the
Hamiltonian (\ref{EQ6}) can be diagonalized for $|\delta|>\gamma$
by solving the eigenvalue problem (\ref{EQ4}). Setting $\sigma =
\delta/\gamma$ one obtains the diagonalizing transformation
\begin{equation}
U = \left(\begin{array}{cc} u_1e^{i\theta_1} & 0 \\
0 & u_1e^{i\theta_2} \end{array}\right), \quad V = \left(\begin{array}{cc} 0 & u_2e^{i\theta_2}  \\
 u_2e^{i\theta_1} & 0 \end{array}\right),
\label{AD1}\end{equation}
with $\quad u_1 = \left(1 - [\sqrt{\sigma^2-1} - \sigma]^2\right)^{-1/2}$ and  $u_2
= [\sqrt{\sigma^2-1} - \sigma]u_1$  (without loss of generality,  the arbitrary
phases $\theta_{1,2}$  can be set to zero). In particular, the new boson operators
read
\begin{equation}
\xi_1 = e^{-i\theta_1}(u_1 a_1 - u_2 a_2^\dag),\quad \xi_2 = e^{-i\theta_2}(u_1 a_2
- u_2 a_1^\dag)
\label{AD2}\end{equation}
and the Hamiltonian reduces to
\begin{equation}
H = (\sqrt{\delta^2-\gamma^2} + \omega)\xi_1^\dag\xi_1 + (\sqrt{\delta^2-\gamma^2}
- \omega)\xi_2^\dag\xi_2 + \sqrt{\delta^2-\gamma^2} - \delta.
\label{AD3}\end{equation}

When   $\sigma = \delta/\gamma$ is time-independent  the diagonalization decouples
the Heisenberg evolution equations for the boson operators $\xi_1$ and $\xi_2$:
\begin{equation}
i\frac{\mathrm{d} \xi_1}{\mathrm{d} t} = (\sqrt{\delta^2-\gamma^2} + \omega)\xi_1,
\quad i\frac{\mathrm{d} \xi_2}{\mathrm{d} t} = (\sqrt{\delta^2-\gamma^2} -
\omega)\xi_2.
\label{AD4}\end{equation}
Otherwise, there is no point in diagonalizing the Hamiltonian  since the Bogoliubov
transformation will be time-dependent and the evolution of the new boson operators
$\xi_1$ and $\xi_2$ will be coupled (similar as the evolution equations for the
original operators $a_1$ and $a_2$).

In the classical stability theory the Hessian  (\ref{EQ7})
describes a collision of two eigenfrequencies  at
$|\delta|=\gamma$ ($\delta \to -\gamma$ from below  or $\delta\to
\gamma$ from above) with the appearance of complex eigenvalues
(eigenfrequencies). This corresponds to the fact that the boson
Hamiltonian (\ref{EQ6}) cannot be diagonalized for
$|\delta|<\gamma$. The corresponding quantum instability can be
thought as the exponential growth of the number of bosons,
described by the boson operators $a_1$ and $a_2$, since the total
number is not conserved.

Though the  Hamiltonian  cannot be diagonalized in the region of instability, the
Bogoliubov transformation (\ref{EQ2}) plays an essential role in the solution of
the Heisenberg evolution equations.  This statement is valid for \textit{any}
quadratic Hamiltonian (\ref{EQ1}) and follows from the fact that the action of the
corresponding time evolution operator on the boson creation and annihilation
operators can be represented by an appropriate time-dependent Bogoliubov
transformation (which is different from the diagonalizing transformation). The
latter fact is just a consequence of the boson commutation relations for the
creation and annihilation operators.

Let us consider the Hamiltonian (\ref{EQ6}) as an illustrative example.  First one
notes that $\omega$ can be eliminated  by the unitary transformation with the
generator $a_1^\dg a_1 - a_2^\dg a_2$, which results in the substitution $a_1 \to
e^{i\omega t}a_1$ and $a_2 \to e^{-i\omega t}a_2$ . Then, taking into account the
symmetry of the transposition $a_1 \leftrightarrow a_2$, the Bogoliubov
transformation solving the Heisenberg evolution equations can be cast as
\begin{equation}
\left(\begin{array}{c}a \\ a^\dg\end{array}\right)
=\left(\begin{array}{cc} U^\dg  & -V^\dg  \\
-\tilde{V}  & \tilde{U}  \end{array}\right)\left(\begin{array}{c}a_0 \\
a_0^\dg\end{array}\right), \quad U = \left(\begin{array}{cc}{u}^*_1 & 0 \\
0 & {u}^*_1\end{array}\right),\quad V = \left(\begin{array}{cc}0 & -{u}^*_2 \\
-{u}^*_2 & 0 \end{array}\right).
\label{EQ8}\end{equation}
The condition $U^\dg  U  - V^\dg  V  = I$ requires that $|u_1|^2 - |u_2|^2 = 1$,
while the second condition from equation (\ref{EQ2}) is satisfied identically.
Differentiating equation (\ref{EQ8}) with respect to time and expressing the
constant operators $a_0$ and $a_0^\dg$ through the time-dependent  operators $a$
and $a^\dg$ by using the inverse Bogoliubov transformation one arrives at the
following linear system for the parameters of the transformation
\begin{equation}
i \frac{\mathrm{d}}{\mathrm{d}t}\left(\begin{array}{c}u_1 \\
{u}^*_2\end{array}\right) = \left(\begin{array}{cc}\delta(t) & \gamma \\
-\gamma & -\delta(t) \end{array}\right)\left(\begin{array}{c}u_1 \\
{u}^*_2\end{array}\right).
\label{EQ9}\end{equation}
A solution of system (\ref{EQ9}) determines the corresponding
solution to the Heisenberg  equations for the boson operators $a$
and $a^\dg$. In fact, the latter equations can be obtained by the
formal replacement: $u_1\to a_1$ and $u^*_2\to a^\dg_2$.  Note
that the  transformation defined by the solution of equation
(\ref{EQ9}) has nothing to do with the diagonalization of the
Hamiltonian.

System (\ref{EQ9}) formally coincides with the normalized classical system, which
obtains from the Hamiltonian (\ref{EQ6}) by replacing the boson operators with
$c$-numbers (equations (\ref{EQ18}) and  (\ref{EQ19}) with $g=0$). The Hamiltonian
(\ref{EQ6}) can also be cast as a linear combination of generators of the group
$SU(1,1)$ (in our case: $\mathcal{K}_0 = (a^\dag_1 a_1 + a^\dag_2 a_2 + 1)/2$,
$\mathcal{K}_+ = a^\dag_1a^\dag_2$ and $\mathcal{K}_- = a_1a_2$ with
$[\mathcal{K}_0,\mathcal{K}_\pm] = \pm\mathcal{K}_\pm$ and
$[\mathcal{K}_+,\mathcal{K}_-] = -2\mathcal{K}_0$), thus the quantum instability
case is formally similar to the problem of parametrically  forced  quantum
oscillator \cite{BZP1}. However, the physics is quite different. For instance,  in
the quantum instability case there no adiabatic frequencies in the region of
instability. The time-dependent Bogoliubov transformation was first used in the
solution of the forced oscillator problem in Ref. \cite{BZP2}.

System (\ref{EQ9}) is equivalent to the approach of Refs.
\cite{Anglin,QCurveCross2}. Let us show, for instance, how the
asymptotic formula of Ref. \cite{Anglin}, relating the boson
operators at $t\to \pm \infty$ in the case $\delta(t) =
\varkappa^2t$, appears in our approach. The first order system
(\ref{EQ9}) is equivalent to a second order equation (satisfied by
both $u_1$ and $u_2$). Setting $t = z/\alpha$ with $\alpha^2 =
2i\varkappa^2$ we get the equation for the parabolic cylinder
(Weber) functions: \cite{SpecFun}
\begin{equation}
\frac{\mathrm{d}^2u_k(z/\alpha)}{\mathrm{d}z^2} + \left[-\frac{z^2}{4} + \nu +
\frac{1}{2}\right]u_k(z/\alpha) = 0
\label{EQ10}\end{equation}
with $\nu = i\gamma^2/2\varkappa^2$.  Using the simplified version
of the integrating Bogoliubov transformation
\begin{equation}
\left(\begin{array}{c}a_1 \\ a_2^\dg\end{array}\right)
=\left(\begin{array}{cc} u_1  & u_2  \\
{u}^*_2  & {u}^*_1  \end{array}\right)\left(\begin{array}{c}a_{10} \\
a_{20}^\dg\end{array}\right),
\label{EQ11}\end{equation}
the asymptotic conditions $u_1 \to 1$ and $u_2 \to 0$ as $ze^{-i\pi/4}\to -\infty$
complemented with the asymptotics of the first derivative derived from equation
(\ref{EQ9}) and the standard asymptotic formulae for the parabolic cylinder
functions one can arrive at the asymptotic formula of Ref. \cite{Anglin}:
\[
a_k(+\infty) = e^{\pi\gamma^2/2\varkappa^2}a_k(-\infty) +
e^{i\eta}[e^{\pi\gamma^2/\varkappa^2} -
1]^{1/2}a^\dg_{k^\prime}(-\infty),
\]
\begin{equation}
 e^{i\eta} \equiv 2^{-i\gamma^2/2\varkappa^2}e^{-i\pi/4}
\left[\frac{\Gamma(1-i\gamma^2/2\varkappa^2)}{\Gamma(1+i\gamma^2/2\varkappa^2)}\right]^{1/2},
\label{EQ12}\end{equation}
with $k^\prime = 2$ for $k=1$ and $k^\prime = 1$ for $k=2$.

It is noted   that  formula (\ref{EQ12}) resembles the well-known
Landau-Zener formula \cite{LZ} (where the Weber functions also
play the key role, see for instance Ref. \cite{LZ1}). However, the
quantum instability due to a  collision of two eigenfrequencies is
completely different from the Landau-Zener problem. This is
manifested by our system (\ref{EQ9}). Indeed, system (\ref{EQ9})
resembles the Landau-Zener system but the evolution matrix on the
r.h.s. of equation (\ref{EQ9}) is not Hermitian and the
normalization condition for the ``wave vector'' $(u_1,u_2)$ is
formulated in the Minkowsky space rather than in the Euclidian
one. One consequence is that, as distinct from the Landau-Zener
case, there is no equivalent of the adiabatic transition for slow
time dependence of $\sigma=\delta/\gamma$, since the boson system
passes through a region ($|\delta|<\gamma$) where the diagonalized
Hamiltonian (\ref{AD3}) is invalid. In other words, there is no
adiabatic eigenfrequencies slowly depending on time in the region
of $|\delta|<\gamma$, instead the system is hyperbolic with the
solution given by a combination of the  exponents
$e^{\pm\sqrt{\gamma^2-\delta^2}t}$.

Therefore, the principal physical difference from the Landau-Zener problem lies in
the fact that, due to the very instability that system (\ref{EQ9}) describes, the
number of  bosons   grows in time and their interaction cannot be neglected at all
times, in general. Let us consider what effect the  interactions may have on the
instability rate. Remembering that the Hamiltonian (\ref{EQ6}) appears as an
approximation to the full nonlinear Hamiltonian (for instance, with the binary
atomic collision term  in the case of molecular dissociation) and eliminating the
$\omega$ by $a_1\to a_1e^{i\omega t}$ and $a_2\to a_2e^{-i\omega t}$ one arrives at
the simplest non-quadratic Hamiltonian describing the instability and interactions
\begin{equation}
H =  \delta(t)(a_1^\dg a_1 + a_2^\dg a_2) + \gamma(a_1^\dg a_2^\dg
+ a_1a_2) + g_{11}(a_1^\dg a_1)^2 + g_{12}a_1^\dg a_1a_2^\dg a_2 +
g_{22}(a_2^\dg a_2)^2.
\label{EQ13}\end{equation}
Below, for simplicity, we will set all the interaction
coefficients equal, $g \equiv g_{11} = g_{12} = g_{22}$. The
Heisenberg evolution equations written for the operators
$\hat{a}_1 = e^{-igKt}a_1$ and $\hat{a}_2^\dg = e^{-igKt}a_2^\dg$,
with $K = a^\dg_1a_1 - a^\dg_2a_2$, read
\begin{equation}
i \frac{\mathrm{d}}{\mathrm{d}t}\left(\begin{array}{c}\hat{a}_1 \\
\hat{a}_2^\dg\end{array}\right)
=\left(\begin{array}{cc} \delta(t) + g +3g\hat{n}_1 & \gamma \\
-\gamma  & -\delta(t)- g -3g\hat{n}_2
\end{array}\right)\left(\begin{array}{c}\hat{a}_1 \\
\hat{a}_2^\dg\end{array}\right),
\label{EQ14}\end{equation}
where $\hat{n}_1 = \hat{a}_1^\dg \hat{a}_1$ and $\hat{n}_2 =
\hat{a}_2^\dg \hat{a}_2$.

System (\ref{EQ14})  is  equivalent to an infinite-dimensional linear system of
equations arising in the Fock representation of the Schr\"odinger equation with the
Hamiltonian (\ref{EQ13}). It turns out that a satisfactory numerical simulation of
such a system requires a large number of the Fock modes to be used (about 10 000)
due to the  spreading of the excitations over a large number of the Fock
amplitudes. This is in contrast to the linear case ($g=0$) when the corresponding
infinite dimensional system in the Fock space is equivalent to the
finite-dimensional system (\ref{EQ9}) for the parameters of the the Bogoliubov
transformation.

Therefore, in the nonlinear case one has to rely on an approximate system. Our
approximation is based on the replacement of the number operators $\hat{n}_1$ and
$\hat{n}_2$ in equation (\ref{EQ14}) by their averages.  Then the evolution of the
boson operators $\hat{a}_1$ and $\hat{a}_2^\dg$ must be given by an appropriate
Bogoliubov transformation. The parameters of the corresponding Bogoliubov
transformation satisfy the system
\begin{equation}
i \frac{\mathrm{d}}{\mathrm{d}t}\left(\begin{array}{c}\hat{u}_1 \\
\hat{u}^*_2\end{array}\right) = \left(\begin{array}{cc}
\delta(t)+g +3g\langle \hat{n}_1\rangle & \gamma \\
-\gamma & -\delta(t) -g  -3g\langle \hat{n}_2\rangle
\end{array}\right)
\left(\begin{array}{c}\hat{u}_1 \\
\hat{u}^*_2\end{array}\right).
\label{EQ15}\end{equation}
Equation (\ref{EQ11}) (with all $a$ replaced by $\hat{a}$)
provides the expression for the boson operators  through their
initial values. Using it one arrives at the following expressions
for the averages:
\begin{equation}
\langle \hat{n}_1\rangle = |\hat{u}_1|^2\langle
\hat{n}_{10}\rangle + |\hat{u}_2|^2(\langle \hat{n}_{20}\rangle
+1), \quad \langle \hat{n}_2\rangle = |\hat{u}_1|^2\langle
\hat{n}_{20}\rangle + |\hat{u}_2|^2(\langle \hat{n}_{10}\rangle
+1)
\label{EQ16}\end{equation}
under the assumption that there is no initial pairing of the
bosons: $\langle \hat{a}_{10}\hat{a}_{20}\rangle = 0$. Using the
normalization condition $|\hat{u}_1|^2 - |\hat{u}_2|^2 = 1$ and
making the phase transformation $\hat{u}_1 = e^{3ig(\langle
\hat{n}_{20}\rangle+1/2)t}u_1^\prime$ and $\hat{u}_2 =
e^{3ig(\langle \hat{n}_{20}\rangle+1/2)t}u_2^\prime$ we arrive at
the  nonlinear system
\begin{equation}
i \frac{\mathrm{d}}{\mathrm{d}t}\left(\begin{array}{c}u^\prime_1 \\
{u^\prime}^*_2\end{array}\right) = \left(\begin{array}{cc}
\delta(t)+g/2 +3g(N_0+1)|u^\prime_1|^2& \gamma \\
-\gamma & -\delta(t) -g/2  -3g(N_0+1)|u^\prime_2|^2
\end{array}\right)
\left(\begin{array}{c}u^\prime_1 \\
{u^\prime}^*_2\end{array}\right),
\label{EQ17}\end{equation}
where $N_0 = \langle \hat{n}_{10}\rangle + \langle
\hat{n}_{20}\rangle $.

At this point it is interesting to compare the approximate system (\ref{EQ17}) with
the classical system resulting from the Hamiltonian (\ref{EQ13}) when the boson
operators are replaced by the $c$-numbers:
\begin{equation}
H_{\mathrm{cl}} = \delta(t)[C^*_1C_1 +C^*_2C_2] +\gamma[C^*_1C^*_2
+ C_1C_2] + g[(C^*_1)^2C_1^2 + C^*_1C_1C^*_2C_2 + (C^*_2)^2C_2^2].
\label{EQ18}\end{equation}
The corresponding classical system reads
\begin{equation}
i \frac{\mathrm{d}}{\mathrm{d}t}\left(\begin{array}{c}C^\prime_1 \\
{C^\prime}^*_2\end{array}\right) = \left(\begin{array}{cc}
\delta(t) +3gK_0|C^\prime_1|^2& \gamma \\
-\gamma & -\delta(t) -3gK_0|C^\prime_2|^2
\end{array}\right)
\left(\begin{array}{c}C^\prime_1 \\
{C^\prime}^*_2\end{array}\right),
\label{EQ19}\end{equation}
where $C_1 = e^{igK_0t}\sqrt{K_0}C_1^\prime$, ${C}^*_2 =
e^{igK_0t}\sqrt{K_0}{C^\prime}^*_2$, with $K_0 = |C_1|^2 - |C_2|^2$ being an
integral of motion. Here $|C_1^\prime|^2 - |C_2^\prime|^2 = 1$.

Systems (\ref{EQ19}) and  (\ref{EQ17}) are similar. One difference between them
lies in the definition of the nonlinearity coefficient. In the classical system it
is proportional to the integral of motion, while in system (\ref{EQ17}) the
nonlinearity coefficient is determined by the initial average of the total number
of the bosons.  (Note that for $K_0=0$ the correspondence between systems
(\ref{EQ19}) and (\ref{EQ17}) cannot be established.) Another difference between
the two systems lies in the physical sense of the variables. In the classical
system $C^\prime_k$ are the (normalized) amplitudes of the perturbation about the
mean-filed state, while in the case of system (\ref{EQ17}) $u_k$ are parameters of
the Bogoliubov transformation which describes the evolution of the boson operators.

\begin{figure}[ht]
\begin{center}
\includegraphics{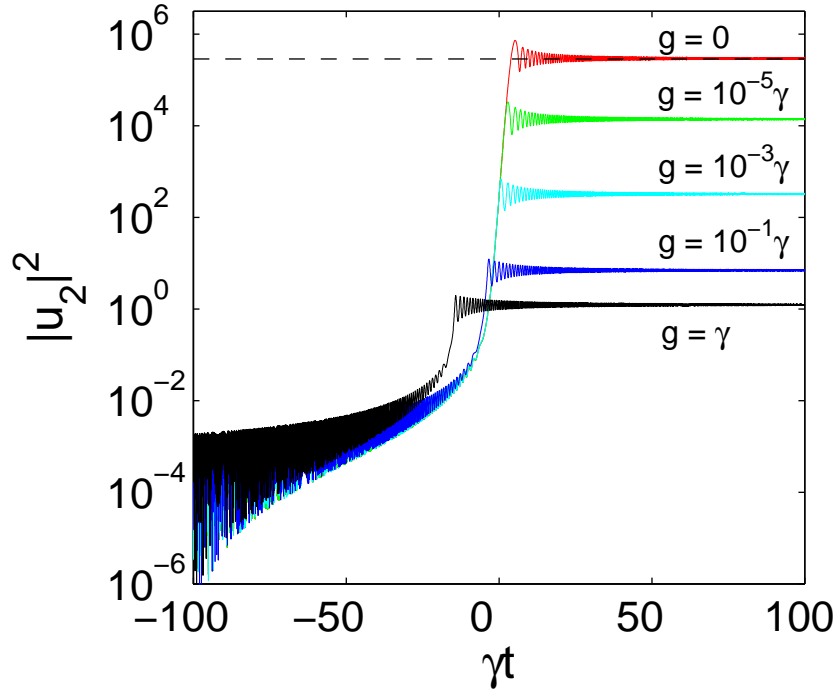}
\caption{\label{FG1}  Numerical solution of the reduced system
(\ref{EQ17}) in the case of repulsive  interactions between the
bosons. Here $\varkappa = 0.5\gamma$. The dashed line is the
asymptotic value of $\langle \hat{n}_1\rangle = |u_2|^2$ for
$g=0$, according to formula (\ref{EQ12}). }
\end{center}
\end{figure}

Consider now the effect of nonlinearity on the quantum
instability. Let us assume that the system is driven through the
instability linearly, $\delta = \varkappa^2 t$. Consider, for
example,  evolution of the system which was initially in the
vacuum state (i.e.  $N_0 = \langle\hat{n}_{10}\rangle +
\langle\hat{n}_{20}\rangle = 0$) and far from resonance:  $-t_0
\gg \gamma/\varkappa^2$.

We are interested in the average values of the number operators, which, according
to formula (\ref{EQ16}), in the case of $N_0=0$ read $\langle\hat{n}_{1}\rangle =
\langle\hat{n}_{2}\rangle = |u_2|^2$.

We have found that in the case of the repulsive interactions between the bosons,
$g>0$, the instability rate is significantly reduced and the asymptotic averages
are lower by orders of magnitude, see fig \ref{FG1}. Note that even a weak
repulsion has a dramatic effect on the resulting asymptotic number of the bosons.

\begin{figure}[ht]
\begin{center}
\includegraphics{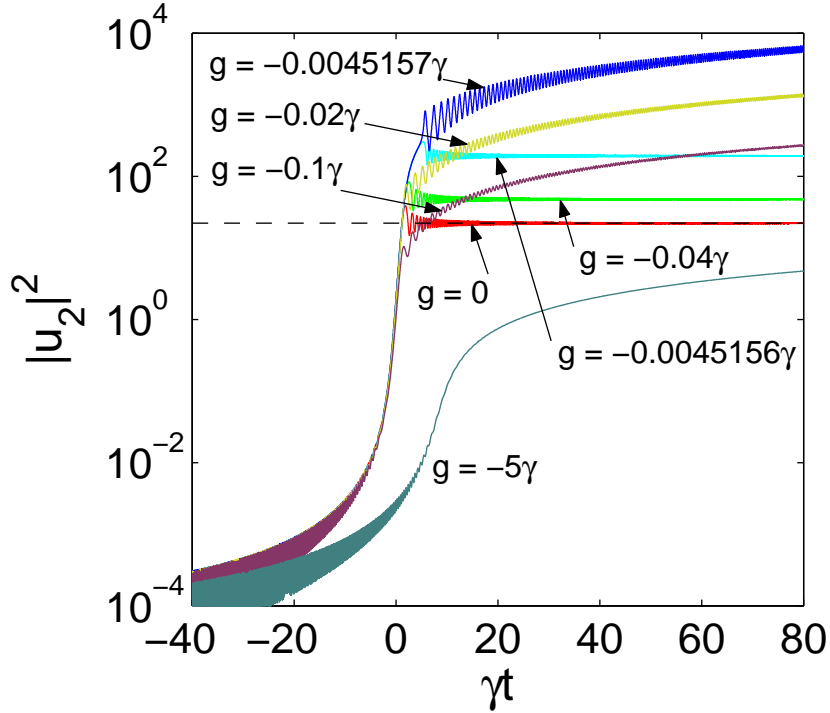}
\caption{\label{FG2}  Numerical solution of the reduced system (\ref{EQ17}) in the
case of attractive interactions between the bosons. Here $\varkappa = \gamma$. The
dashed line is the corresponding asymptotic result  for $g=0$. We have used $t_0 =
-100$. }
\end{center}
\end{figure}

In the case of the attractive interactions, $g<0$, the effect of a
weak attraction consists in enhancing the production of the
bosons, as seen from fig. \ref{FG2}. On the other hand, a strong
attractive interaction decreases the instability rate.

In conclusion, it is shown that the quantum instability of the mean-field theory
for identical bosons can be described by an appropriate Bogoliubov transformation.
The relation to the instability in the corresponding classical system is
established. We argue that a quantum instability in a system of identical bosons is
strongly affected by the interactions. In the case of  repulsion or strong
attraction between the bosons the instability rate is suppressed. On the other
hand, a weak attraction significantly enhances the instability rate. These results
can have applications in the field of Bose-Einstein condensates of quantum gases.

\section*{Acknowledgements}
This work was supported by the CNPq-FAPEAL grant of Brazil. The author would like
 thank Professor A.M. Kamchatnov for invaluable discussion.

\end{document}